\documentclass[prd,aps,10pt,twocolumn,nofootinbib,superscriptaddress,showpacs]{revtex4-2}

\usepackage{amssymb,amsmath}
\usepackage{xcolor}
\usepackage[utf8]{inputenc}

\usepackage{graphicx}

\usepackage[makeroom]{cancel}
\usepackage[caption=false]{subfig}
\usepackage[colorlinks=true,
            citecolor=green,
            linkcolor=red,
            filecolor=cyan,
            urlcolor=magenta,
            backref=false]{hyperref}

\newcommand{\dd}{\mbox{d}}

\newcommand{\ts}[1]{{\boldsymbol{#1}}}

\begin{document}

\title{What happens to topological invariants (and black holes) in singularity-free theories?}
\author{Jens Boos}
\email{jens.boos@kit.edu}
\affiliation{Institute for Theoretical Physics, Karlsruhe Institute of Technology, D-76128 Karlsruhe, Germany}

\date{November 18, 2024}

\begin{abstract}
Potentials arising in ultraviolet-completed field theories can be devoid of singularities, and hence render spacetimes simply connected. This challenges the notion of topological invariants considered in such scenarios. We explore the classical implications for (i) electrodynamics in flat spacetime, (ii) ultrarelativistic gyratonic solutions of weak-field gravity, and (iii)the Reissner--Nordstr\"om black hole in general relativity. In linear theories, regularity spoils the character of topological invariants and leads to radius-dependent Aharonov--Bohm phases, which are potentially observable for large winding numbers. In general relativity, the physics is richer: The electromagnetic field can be regular and maintain its usual topological invariants, and the resulting geometry can be interpreted as a Reissner--Nordstr\"om black hole with a spacetime region of coordinate radius $\sim q^2/(GM)$ cut out. This guarantees the regularity of linear and quadratic curvature invariants ($\mathcal{R}$ and $\mathcal{R}^2$), but does not resolve singularities in invariants such as $\mathcal{R}^p\Box^n \mathcal{R}^q$, reflected by conical or solid angle defects. This motivates that gravitational models beyond general relativity need to be considered. These connections between regularity (= UV properties of field theories) and topological invariants (= IR observables) may hence present an intriguing avenue to search for traces of new physics and identify promising modified gravity theories.
\end{abstract}

\maketitle

\section{Introduction}

The field configurations for point sources in classical field theory typically contain singularities, which can render physical observables such as field strengths or self energies infinite. Physically, this points towards an inconsistency of the theory under consideration, and it is expected that in so-called ultraviolet (UV) complete theories such singularities are absent. However, finding such theories can be quite difficult and one hence often deals instead with effective field theories that regularize singularities with a regulator scale, corresponding to a UV cutoff $\Lambda < \infty$ or a minimal length $\ell > 0$. However, in the absence of a guiding principle for the form of such effective field theories it becomes all the more important to search for reliable and robust experimental signatures of such regulators. Unfortunately, quite often such effects are suppressed strongly by the typical energy of the system ($E/\Lambda \lll 1$) or by the size of it ($\ell/L \lll 1$). 

In this paper, we consider a different type of effect: the deformation of topological invariants by non-singular field theories. If a physical observable can be expressed through topological quantities (such as asymptotic fluxes, more on that below), its value can usually be measured rather robustly, since deformations of the measurement surface or distance do not affect its theoretical value. Conversely, if such a topological relation were to be altered, it could give rise to unique experimental signatures, such as distance-dependent charges or winding number-dependent topological phases. We will report on such violations in the context of classical field theory in a few examples and briefly point out their possible significance for experimental investigations.

For instance, consider the electric field of a point charge in flat spacetime,
\begin{align}
\boldsymbol{F} = \frac{Q}{4\pi\epsilon_0}\frac{1}{r^2} \dd t \wedge \dd r \, .
\end{align}
Its charge is given by the surface integral
\begin{align}
\label{eq:intro-charge}
Q = \epsilon_0 \oint_{S^2_r} \star \boldsymbol{F} = \oint_{S^2_r} \left(\frac{Q}{4\pi}\frac{1}{r^2}\right) \left(r^2 \sin\theta \right) \dd\theta \wedge \dd\varphi \, .
\end{align}
Key to this relation is the precise cancellation between how the field strength \emph{decreases} with distance, and how the surface area of a 2-sphere $S^2_r$ \emph{grows} with radius. In other words, we may think of the charge of a point particle as a topological invariant. Suppose now that we work in a theory of electrodynamics wherein the field of a point charge is non-singular, such that
\begin{align}
\boldsymbol{F}_\text{reg} = \frac{Q}{4\pi\epsilon_0} F_\text{reg}(r) \, \dd t \wedge \dd r \, ,
\end{align}
where $F_\text{reg}$ is some regular function that is finite as $r \rightarrow 0$, and that asymptotically behaves as $F_\text{reg}(r \rightarrow\infty) \approx 1/r^2$. If we were to compute the charge of that configuration, we find instead a radius-dependent quantity,
\begin{align}
Q(r) = \epsilon_0 \oint_{S^2_r} \star \boldsymbol{F}_\text{reg} = Q - \delta Q(r) \not= Q \, .
\end{align}
In such a theory, one only recovers the charge at large distances, $\delta Q(r\rightarrow\infty) = 0$, and its topological nature is lost. If we want to recover a topological notion of charge in such a non-singular theory, we need to move beyond flat spacetime and consider the charge's back-reaction onto the background metric. In other words: the surface element of a sphere around a charge cannot scale with $r^2$, because then no precise cancellation can occur.

Before discussing the curved spacetime treatment, it is instructive to consider a range of examples to motivate the generality and scope of the considered problem. Hence, for the remainder of this paper, we will focus on the following three scenarios:
\begin{itemize}
\item[(i)] Electrodynamics: non-singular field theories alter the topological notion of electric charge and winding number around a solenoid in flat spacetime, and spoil the electromagnetic duality in vacuum.\\[-1.4\baselineskip]
\item[(ii)] Weak-field gravity: ultrarelativistic objects (``gyratons'') are described by topological invariants in the direction transverse to their motion. Again, these charges and winding numbers will lose their topological meaning in non-singular field theories.\\[-1.4\baselineskip]
\item[(iii)] General relativity: Electro-vacuum black hole solutions (extendable to the rotating case with cosmological constant) can feature non-singular electromagnetic fields with properly defined topological charges. This induces regular gravitational linear and quadratic curvature, at the cost of divergences of higher-order invariants.
\end{itemize}
In settings where large winding numbers can be measured accurately or where field fluxes are measured as functions of distance, these scenarios hence provide avenues to test UV physics with IR observations via the ``smoking gun'' signature of deformed topological invariants.

\subsection{Preliminaries}
Let us first make more precise our notion of ``singularity-free field theories'' in the context of classical field theories as employed in the remainder of this paper. We will largely follow the notation and results of \cite{Edholm:2016hbt,Giacchini:2016xns,Buoninfante:2018xiw,Boos:2018bxf,Giacchini:2018wlf,Boos:2020kgj,Kolar:2020bpo,Boos:2020ccj,Boos:2020qgg,Boos:2021suz} and focus on static Green functions in flat spacetime,
\begin{align}
\dd s^2 = -\dd t^2 + \dd \ts{x}^2 \, , \quad \dd \ts{x}^2 = \dd x^2 + \dd y^2 + \dd z^2 \, ,
\end{align}
and we will denote the radial distance by $r$ for simplicity. Loosely speaking, static Green functions serve as the inverse of elliptic differential operators, where the differential operator is specified by the theory under consideration. Classical field theory in the static limit is typically formulated in terms of the spatial part of the d'Alembert operator,
\begin{align}
\nabla^2 = \partial_x^2 + \partial_y^2 + \partial_z^2 \, .
\end{align}
Singularity-free field theories are then defined by making the substitution
\begin{align}
\nabla^2 \rightarrow f\left( \ell^2 \nabla^2 \right) \nabla^2 \, ,
\end{align}
where $f$ is called a form factor, $\ell > 0$ is a regulator, and one has $f(0) = 1$ such that in the limiting case of $\ell \rightarrow 0$ one recovers standard field theory. The static Green function $G_d(r)$ of standard field theory and the singularity-free Green function $\mathcal{G}_d(r)$ are then defined via
\begin{align}
\nabla^2 G_d(r)& = -\delta{}^{(d)}(\ts{x}) \, , \label{eq:gf} \\
\nabla^2 f(\ell^2 \nabla^2) \mathcal{G}_d(r) &=  -\delta{}^{(d)}(\ts{x}) \, . \label{eq:gf-f}
\end{align}
The latter is solved by the expression
\begin{align}
\mathcal{G}_d(r) = \frac{1}{(2\pi)^{d/2}r^{d-2}} \int\limits_0^\infty \dd z \, \frac{z^{\frac{d-4}{2}}}{f\left(-\frac{z^2}{r^2}\right)} J_{\frac{d}{2}-1}(z) \, ,
\end{align}
where $J$ denotes the Bessel function of the first kind, and one obtains the standard Green function $G_d(r)$ by inserting $f=1$. We note in passing that the above representation implies a recursion relation between Green functions,
\begin{align}
\partial_r G_d &= -2\pi r G_{d+2}(r) \, , \label{eq:gf-rec} \\
\partial_r \mathcal{G}_d &= -2\pi r \mathcal{G}_{d+2}(r) \, , \label{eq:gf-rec-f}
\end{align}
which will be useful whenever we encounter differentiations of Green functions in the following. We are interested in singularity-free field configurations, which for a field $\phi(r)$ we define as follows:
\begin{itemize}
\item $\phi(r\rightarrow 0)$ is finite;\\[-1.4\baselineskip]
\item $\phi'(r\rightarrow 0) = 0$.
\end{itemize}
For the Green function above, close to $r=0$ one finds
\begin{align}
\mathcal{G}_d(r\rightarrow 0) = \frac{1}{(4\pi)^{d/2} \Gamma\left( \frac{d}{2} \right)} \int\limits_0^\infty \dd z \, \frac{z^{\frac{d-4}{2}}}{f\left(-z\right)} + \mathcal{O}(r^2) \, ,
\end{align}
which is a singularity-free quantity if the $z$-integral converges. For $f=1$ (standard field theory) this does not happen, and hence the regularity of fields is dictated by the properties of the function $f$ at large arguments (or, in other words, its UV properties). Hence, if $f$ increases fast enough for large, negative arguments, the theory will feature non-singular Green functions. In order to compare Green functions of standard field theory to those of singularity-free field theory, we write
\begin{align}
\label{eq:deviation}
\mathcal{G}_d(r) = \Delta_d(r) \, G_d(r) \, ,
\end{align}
where we will refer to $\Delta_d(r)$ as a deviation function. Last, it is also possible to express the regularity of Green functions via inverting the differential operator $f(\ell^2\nabla^2)$ symbolically in the defining equation \eqref{eq:gf-f} such that
\begin{align}
\nabla^2 \mathcal{G}_d(r) &= - \, f^{-1}(\ell^2 \nabla^2) \, \delta{}^{(d)}(\ts{x}) \equiv - \, \delta{}^{(d)}_\ell(\ts{x}) \, , \label{eq:gf-f-2}
\end{align}
where $\delta{}^{(d)}_\ell(\ts{x})$ is a so-called nascent $\delta$-function that reduces to the distributional $\delta$-function in the limit \cite{Giacchini:2018wlf,Boos:2021suz}
\begin{align}
\lim\limits_{\ell\rightarrow 0} \delta{}^{(d)}_\ell(\ts{x}) = \delta{}^{(d)}(\ts{x}) \, .
\end{align}
We have not yet specified the form factor $f$ because there is a plethora of possible choices. For simplicity and illustrative purposes, for the remainder of the paper we will adopt the choice 
\begin{align}
f(\ell^2\nabla^2) = \exp\left( -\ell^2\nabla^2 \right) \, . \label{eq:f-choice}
\end{align}
For analytical expressions of the static Green functions for $d=2,3,4,5$ in that case, we refer to appendix \ref{app:gf}.

\section{Electromagnetism}
Consider flat spacetime in spherical coordinates,
\begin{align}
\begin{split}
\label{eq:metric}
\dd s^2 &= -\dd t^2 + \dd r^2 + r^2 \dd\Omega^2 \, , \\
\dd \Omega^2 &= \dd\theta^2 + \sin^2\theta\dd\varphi^2 \, ,
\end{split}
\end{align}
where here and in what follows we will set $c\equiv 1$. We will denote the coordinates as $x{}^\mu = (t,r,\theta,\phi)$ and write collectively $x{}^\mu = (t,\ts{x})$ for the spatial components. We write the Maxwell equations as
\begin{align}
\dd \ts{F} = 0 \, , \quad \dd \star \ts{F} = \ts{j} \, ,
\end{align}
where $\ts{F}$ is the field strength 2-form
\begin{align}
\ts{F} = \frac12 F{}_{\mu\nu} \dd x{}^\mu \wedge \dd x{}^\nu \, ,
\end{align}
the symbol $\ts{j}$ is the current 3-form, and ``$\star$'' denotes the Hodge dual. Expressed in components we have instead
\begin{align}
\begin{split}
& \frac{1}{\sqrt{-g}} \partial{}_\nu \left( \sqrt{-g} F{}^{\mu\nu} \right) = j{}^\mu \, , \\
& \partial{}_\mu F{}_{\nu\rho} + \partial{}_{\nu} F{}_{\rho\mu} + \partial{}_\rho F{}_{\mu\nu} = 0 \, .
\end{split}
\end{align}
Last, the homogeneous equations imply that locally there exists a potential 1-form $\ts{A} = A{}_\mu \dd x{}^\mu$ such that
\begin{align}
\ts{F} = \dd \ts{A} \, , \quad
F{}_{\mu\nu} = \partial{}_\mu A{}_\nu - \partial{}_\nu A{}_\mu \, .
\end{align}

\subsection{Point charge and duality}
\label{sec:monopoles}
In the Lorenz gauge ($\partial{}_\mu A{}^\mu = 0$) the electromagnetic potential of a point charge $q$ satisfies
\begin{align}
\nabla^2 A{}_\mu = -(Q/\epsilon_0) \delta{}^t_\mu \delta{}^{(3)}(\ts{x}) \, .
\end{align}
where $\epsilon_0$ denotes the permittivity of flat spacetime. Using the Green function method, the solution is
\begin{align}
\ts{A} &= (Q/\epsilon_0)G_3(r) \dd t \, , \\
\ts{F} &= \dd \ts{A} = -(Q/\epsilon_0) G_3'(r) \dd t \wedge \dd r \, .
\end{align}
In standard Maxwell theory one has  $G_3(r) = 1/(4\pi r)$, such that the field strength is given by Coulomb's law,
\begin{align}
\ts{F} = \frac{Q}{4\pi\epsilon_0}\frac{1}{r^2} \dd t \wedge \dd r \, .
\end{align}
Inspecting the background metric \eqref{eq:metric} it is clear that the dual of this field strength is a magnetic monopole,
\begin{align}
\star \ts{F} = P \sin\theta \, \dd\theta\wedge\dd\varphi \, , \quad P = \frac{Q}{4\pi\epsilon_0} \, .
\end{align}
Crucial for this duality is the cancellation of the field strength's $1/r^2$-divergence with the $r^2$-area factor from the metric, or, in other words, Gauss' law.

Let us now ask what happens if the potential $\ts{A}(r)$ becomes regular at $r=0$. The previous considerations imply that such a regularity condition imposed on a potential $\ts{A}(r)$ is equivalent to regularity condition on the corresponding Green function, whose singular behavior at $r=0$ in turn tracks to the distributional right-hand side of its defining equation \eqref{eq:gf}. Hence, we substitute $G_3(r) \rightarrow \mathcal{G}_3(r)$, and, after making use of the recursion relation \eqref{eq:gf-rec-f} and the deviation function \eqref{eq:deviation} we find
\begin{align}
\ts{F}_\text{reg} = \frac{Q}{4\pi \epsilon_0}\frac{1}{r^2}\Delta_5(r) \dd t \wedge \dd r \, .
\end{align}
Comparing to Eq.~\eqref{eq:intro-charge} in the Introduction, we find for the charge instead
\begin{align}
\label{eq:q-point-charge}
Q_\text{reg}(r) = \epsilon_0 \oint_{S^2_r} \star \boldsymbol{F}_\text{reg} = Q \Delta_5(r) \, .
\end{align}
This is manifestly radius-dependent. For large distances, however, we demand that $\Delta_5(r \rightarrow \infty) = 1$ such that $Q(r\rightarrow\infty) = Q$. For the dual field strength we find
\begin{align}
\star \ts{F}_\text{reg} = \Delta_5(r) \, P \sin\theta \, \dd\theta \wedge \dd\varphi \, ,
\end{align}
and since $\dd \star \ts{F}_\text{reg} \not= 0$ this does not describe a monopole, spoiling the duality encountered in the Introduction.

\subsection{Solenoid field}
Let us consider the magnetic field of an infinite string that is magnetized the $x^j$-direction (where from now on we denote spatial components via Latin indices such that $\ts{x} = (x^i)$. To that end we introduce the magnetization axial vector $\mu^j$ that is related to the magnetization 2-form via $\mu_{ij} = \epsilon{}_{ijk}\mu^k$. The Maxwell equations for such an object, in Lorenz gauge, take the form
\begin{align}
\nabla^2 A_i = -\mu{}_i{}^k \partial{}_k \delta{}^{(2)}(\ts{x}_\perp) \, ,
\end{align}
where $\ts{x}_\perp$ denotes the directions perpendicular to $\mu^j$, and we denote $|\ts{x}_\perp|^2 = \rho^2$. The solution is given by the following expression:
\begin{align}
\ts{A} = A_i \dd x{}^i = \mu{}_i{}^k \partial{}_k G_2(\rho) \dd x^i \, .
\end{align}
Let us now fix the magnetization along the $z$-direction such that $\mu_{xy} = -\mu_{yx} \equiv \mu$ with all other components vanishing, and define the transverse distance $\rho^2 \equiv x^2 + y^2$. Then one finds the simple expression
\begin{align}
\ts{A} = 2\pi\mu\rho^2 G_4(\rho) \dd\varphi = \frac{\mu}{2\pi} \dd\varphi \, .
\end{align}
Hence, the magnetization is a topological quantity,
\begin{align}
\mu = \oint_{C_\rho} \ts{A} = \int\limits_0^{2\pi} \frac{\mu}{2\pi} \dd\varphi \, ,
\end{align}
where $C_\rho$ denotes a closed path around the $z$-axis at radius $\rho$. By linearity, the non-singular field takes the form
\begin{align}
\ts{A}_\text{reg} = 2\pi\mu\rho^2 \mathcal{G}_4(\rho) \dd\varphi = \Delta_4(\rho) \ts{A} \, .
\end{align}
Explicitly, the deviation function $\Delta_4(\rho)$ takes the form
\begin{align}
\Delta_4(\rho) = 1 - \exp\left[-\rho^2/(4\ell^2)\right] \, .
\end{align}
Hence, in the non-singular theory, we instead find
\begin{align}
\mu_\text{reg}(\rho) = \oint_{C_\rho} \ts{A}_\text{reg} = \mu \, \Delta_4(\rho) \, ,
\end{align}
a manifestly radius-dependent quantity. Again, the price we have to pay for a non-singular field configuration lies in the loss of a topological invariant. For large distances, however, we recover $\mu_\text{reg}(\rho\rightarrow\infty) \approx \mu$. This winding number appears in magnetic Aharonov--Bohm phases and hence leads to path-dependent behavior for finite $\rho$, in stark departure from standard field theory.

\section{Weak-field gravity}
\label{sec:linearized-gravity}
Due to the similarity of electrodynamics and linearized gravity it is not surprising that the conclusions of the previous section can be straightforwardly extended to linearized gravity. For this reason, we will be brief. Moreover, we will take the first steps towards statements valid in general relativity by discussing so-called gyraton metrics. These metrics describe ultrarelativistic objects and, even though obtained in the linearized theory, are solutions of general relativity in four dimensions \cite{Aichelburg:1970dh}. As we will see, their topological invariants behave rather similarly to those of the solenoid field, even though this time the winding number is defined around a null axis.

In this section, we consider small deviations around flat spacetime,
\begin{align}
g{}_{\mu\nu} = \eta{}_{\mu\nu} + h{}_{\mu\nu} \, , 
\end{align}
where $\eta{}_{\mu\nu} = \text{diag}(-1,1,1,1)$ is the metric of flat spacetime expressed in Cartesian coordinates $x{}^\mu = (t,x,y,z)$, and $h{}_{\mu\nu} \ll 1$ captures the gravitational potential. In the harmonic gauge, expressed in terms of the trace-reversed $\hat{h}{}_{\mu\nu}$, the field equations simplify to
\begin{align}
\Box \hat{h}{}_{\mu\nu} = -2\kappa T{}_{\mu\nu} \, ,
\end{align}
where $\kappa = 8\pi G$ is the gravitational constant, $T{}_{\mu\nu}$ is the conserved energy-momentum tensor of an external matter source, and $\Box = -\partial_t^2 + \nabla^2$ denotes the d'Alembert operator of flat spacetime. In case of static sources, which we will consider in this work, the solution is given by
\begin{align}
\hat{h}{}_{\mu\nu} = 2\kappa \int \dd^3 y \, T{}_{\mu\nu}(\ts{y}) G_3(\ts{x} - \ts{y})
\end{align}
Alternatively, defining the objects
\begin{align}
\ts{\hat{h}} = h{}_{\mu\nu} \dd x{}^\mu \dd x{}^\nu \, , \quad \ts{T} = T{}_{\mu\nu} \dd x{}^\mu \dd x{}^\nu \, , 
\end{align}
we can write
\begin{align}
\ts{\hat{h}} = 2\kappa \int \dd^3 y \, \ts{T} (\ts{y}) G_3(\ts{x} - \ts{y}) \, .
\end{align}
Throughout, retardation effects do not play any role due to the staticity of the source.

\subsection{Point particle}
The conserved energy-momentum tensor of a point particle in linearized gravity is given by
\begin{align}
\ts{T} = M \, \delta{}^{(3)}(\ts{x}) \dd t^2 \, .
\end{align}
Making the static and spherically symmetric ansatz
\begin{align}
\ts{h} = \phi(r) \dd t^2 + \psi(r)\dd \ts{x}^2 \, ,
\end{align}
we find for the gravitational field
\begin{align}
\phi(r) = \psi(r) = \kappa M G_3(r) = \frac{2GM}{r} \, ,
\end{align}
resulting in the well-known linearized Schwarzschild metric in Cartesian coordinates,
\begin{align}
\ts{g} = -\left(1-\frac{2GM}{r}\right)\dd t^2 + \left(1+\frac{2GM}{r}\right)\dd \ts{x}^2 \, .
\end{align}
Noticing that $\xi = \partial_t$ is a timelike Killing vector of this metric, we can define the Killing 1-form $\ts{\xi} = \xi{}_\mu \dd x{}^\mu$ and extract the Komar mass $M_\text{K}$ via
\begin{align}
\begin{split}
G M_\text{K} &= \frac{1}{8\pi} \oint_{S^2_r} \star \dd \ts{\xi} = \frac{1}{8\pi} \oint_{S^2_r} \star \left( \frac{2GM}{r^2} \dd t \wedge \dd r \right) \\
&= \frac{GM}{4\pi} \oint_{S^2_r} \sin^2\theta\,\dd\theta \wedge \dd\varphi = GM \, .
\end{split}
\end{align}
However, in the non-singular theory one again works instead with the regularized Green function $\mathcal{G}_3(r)$ such that one then obtains \cite{Boos:2020qgg}
\begin{align}
\phi_\text{reg}(r) = \psi_\text{reg}(r) = \kappa M \mathcal{G}_3(r) = \frac{2GM}{r} \Delta_3(r) \, ,
\end{align}
Consequently, the Komar mass becomes
\begin{align}
G M_\text{K,\,reg} = GM \Delta_5(r) \, .
\end{align}
Similar to the case of the electric point charge, cf.~Eq.~\eqref{eq:q-point-charge}, the Komar mass is now radius-dependent, which is unexpected within static and spherically symmetric vacuum configurations in linearized gravity.

\subsection{Rotating string}
As an analog to the solenoid field, let us now consider the gravitational field of an infinitely extended rotating string. Assuming that it spans the entire $z$-axis, and that its angular momentum is also aligned along the $z$-axis, its energy-momentum tensor is \cite{Boos:2020kgj,Kolar:2020bpo}
\begin{align}
\begin{split}
\ts{T} &= \mu_s ( \dd t^2 - \dd z^2) \delta(x)\delta(y) \\
&\hspace{11pt}+ 2j_s \left[  \delta(x) \delta'(y) \, \dd x - \delta'(x)\delta(y) \, \dd y \right] \dd t \, .
\end{split}
\end{align}
Here, $\mu_s$ is the string tension, $j_s$ is the string angular momentum, and $\delta'$ denotes the derivative of the delta function, which is only well-defined inside an integral. Making the ansatz 
\begin{align}
\ts{h} = \phi( \dd x^2 + \dd y^2 ) + 2\ts{A} \dd t \, ,
\end{align}
and introducing polar coordinates according to $\rho^2 = x^2+y^2$ such that $x = \rho\cos\varphi$ and $y=\rho\sin\varphi$, we find
\begin{align}
\phi &= 16\pi G \mu_s G_2(\rho) \, , \\
\ts{A} &= 16\pi^2 G j_s \rho^2 G_4(\rho) \, \dd\varphi = 4G j_s \dd\varphi \, .
\end{align}
This field configuration is characterized by two topological invariants: the angle deficit (stemming from the string tension) and the gravitomagnetic charge (stemming from the string rotation). The angle deficit $\delta\varphi$ can be evaluated by considering the ratio between the circumference of a circle of constant radius $\rho$ and the proper radius corresponding to the radius $\rho$,
\begin{align}
\delta\varphi &= 2\pi - \frac{C(\rho)}{R(\rho)} \, , \\
C(\rho) &= \int_0^{2\pi} \dd\varphi \, g_{\phi\phi} = 2\pi \rho \sqrt{1 + \phi(\rho)} \, , \\
R(\rho) &= \int_0^\rho \dd\bar{\rho} \, g_{\rho\rho} = \int_0^\rho \dd\bar{\rho} \sqrt{1 + \phi(\bar{\rho})} \, .
\end{align}
For small string tension, $\mu_s G \ll 1$, one finds
\begin{align}
\label{eq:angle-deficit}
\delta\varphi &= -16\pi^2 G \mu_s \bigg[ G_2(\rho) - \frac{1}{\rho} \int_0^\rho \dd\bar{\rho} \, G_2(\bar{\rho}) \bigg] \, .
\end{align}
In linearized gravity one has $G_2(r) = -\log(r/r_0)/(2\pi)$ and hence the angle deficit becomes
\begin{align}
\delta\phi = 8\pi G \mu_s \, .
\end{align}
Inspecting Eq.~\eqref{eq:angle-deficit}, it is not obvious why in linearized gravity the angle deficit is a constant---and, since it does not depend on the radius, why it is a topological quantity. However, one may show that the quantity in brackets in Eq.~\eqref{eq:angle-deficit} is a constant $\alpha$ if and only if
\begin{align}
G_2(\rho) = \alpha \log \rho + \text{const.} \, ,
\end{align}
which is the case for the linearized limit of general relativity. However, for singularity-free field theories one again needs to replace $G_2(\rho) \rightarrow \mathcal{G}_2(\rho) = G_2(\rho) \Delta_2(\rho)$, and hence the angle deficit becomes radius-dependent, $\delta\varphi_\text{reg} = \delta\varphi_\text{reg}(\rho)$. Due to the radial integration, however, we are not aware of any closed-form expression of \eqref{eq:angle-deficit} in terms of the deviation functions $\Delta_d(\rho)$. For our example case \eqref{eq:f-choice}, however, it is straightforward to compute
\begin{align}
\begin{split}
\delta\varphi_\text{reg}(\rho) &= 8\pi G \mu_s \left[ 1 - \frac{\sqrt{\pi}\ell}{\rho} \text{erf}\left( \frac{\rho}{2\ell} \right) \right] \\
&\overset{*} = 8\pi G \mu_s \left[ 1 - 4\pi\sqrt{\pi} \, \ell \, G_3(\rho)\, \Delta_3(\rho) \right] \, ,
\end{split}
\end{align}
where the asterisk over the last equality denotes that this equality is only valid for our special example case of a non-singular field theory of Eq.~\eqref{eq:f-choice}. In the limit of $\rho\rightarrow\infty$, we have $\Delta_3(\rho \rightarrow \infty) = 1$ and $G_r(\rho\rightarrow\infty) = 0$ such that $\delta\varphi_\text{reg}(\rho\rightarrow\infty) = 8\pi G \mu_s$ \cite{Boos:2020kgj}.

Second, it is possible to define a gravitomagnetic field $\ts{B} = \dd \ts{A}$. Since $\ts{A} = 4G j_s \dd\varphi$, one might be tempted to conclude $\ts{B} = 0$, but keeping in mind that the variable $\varphi$ is angular we notice that $\ts{A}$ is not globally exact. For this reason, $\ts{B}$ is singular on the $z$-axis. Hence, the string angular momentum is a topological invariant,
\begin{align}
j_s = \frac{1}{8\pi G} \oint_{C_\rho} \ts{A} = \int\limits_0^{2\pi} \frac{j_s}{2\pi} \dd\varphi \, ,
\end{align}
where $C_\rho$ describes one circle of radius $\rho$. In the non-singular field theory case, however, one simply finds
\begin{align}
\ts{A}_\text{reg} = 4 G j_s \Delta_4(\rho)
\end{align}
and hence
\begin{align}
j_\text{s,\,reg}(\rho) = \frac{1}{8\pi G}\oint_{C_\rho} \ts{A} = j_s \Delta_4(\rho) \, .
\end{align}
Again, we arrive at a radius-dependent quantity.

\subsection{Ultrarelativistic gyratons}
As a last step in linearized theory, let us briefly explore how some of the previously explored notions of topological invariants can be carried over to the manifestly time-dependent case. For that purpose, it is instructive to consider the topological meaning of the magnetization of an ultrarelativistic source \cite{Boos:2020ccj}. We parametrize flat spacetime in the rest frame of that source as follows:
\begin{align}
\dd s^2 = -\dd \bar{t}^2 + \dd \bar{\xi}^2 + \dd \ts{x}_\perp^2 \, ,
\end{align}
where we singled out one spatial direction $\bar{\xi}$ and split off its orthogonal components in $(x{}^\alpha) = \ts{x}_\perp$. All barred quantities are tied to that reference frame, wherein the source takes the form of an infinitely extended string with angular momentum that may vary along its axis,
\begin{align}
\ts{T} = \left\{ \bar{\lambda}(\bar{\xi}) \, \dd t^2 + 2\left[ \partial_\alpha \bar{j}{}_\beta{}^\alpha(\bar{\xi}) \right] \dd t \, \dd x{}^\beta \right\} \, \delta^{(2)}(\ts{x}_\perp) \, .
\end{align}
Here, $\bar{\lambda}(\bar{\xi})$ is a mass line density, and $\bar{j}{}_{ij}(\bar{\xi})$ is the angular momentum line density. The gravitational field of such a source is then
\begin{align}
\ts{h} &= \bar{\phi} \left( \dd \bar{t}^2 + \dd \bar{\xi}^2 + \dd \ts{x}_\perp^2 \right) + 2 \ts{A}\dd \bar{t} \, , \\
\bar{\phi} &= \kappa \int \dd\bar{\xi}' \bar{\lambda}(\bar{\xi}') G_3(\bar{r}) \, , \\
\ts{A} &= 2\pi\kappa \int \dd\bar{\xi}' \bar{j}{}_{ij}(\bar{\xi}') x{}^i_\perp \dd x{}_\perp^j G_5(\bar{r}) \, , \\
\bar{r}^2 &= (\bar{\xi} - \bar{\xi}')^2 + \ts{x}_\perp^2
\end{align}
To arrive at the metric of an ultrarelativistic source, let us now boost this metric via
\begin{align}
\bar{t} = \gamma(t-\beta \xi) \, , \quad \bar{\xi} = \gamma(\xi - \beta t) \, , \\
u = \frac{1}{\sqrt{2}}(t-\xi) \, , \quad v = \frac{1}{\sqrt{2}}( t + \xi ) \, .
\end{align}
We are interested in taking the limit of $\beta \rightarrow 1$, while keeping the product of the gamma parameter $\gamma = 1/\sqrt{1-\beta^2}$ with the mass density and angular momentum density, respectively, fixed. This is also called the \emph{Penrose limit} \cite{Penrose:1976} and we define it as
\begin{align}
\lambda(u) &= \lim\limits_{\gamma\rightarrow\infty} \sqrt{2}\gamma^2\bar{\lambda}(-\sqrt{2}\gamma u) \, , \\
j_{ij}(u) &= \lim\limits_{\gamma\rightarrow\infty} \sqrt{2}\gamma \bar{j}_{ij}(-\sqrt{2}\gamma u) \, .
\end{align}
This limit guarantees the following relations:
\begin{align}
\int\limits_{-\infty}^\infty \dd \bar{\xi} \, \bar{\lambda}(\bar{\xi}) &= \int\limits_{-\infty}^\infty \dd u \, \lambda(u) \, , \\
\int\limits_{-\infty}^\infty \dd \bar{\xi} \, \bar{j}_{ij}(\bar{\xi}) &= \int\limits_{-\infty}^\infty \dd u \, j_{ij}(u) \, .
\end{align}
In this scaling limit we make use of the following identity for static Green functions (also valid for $\mathcal{G}_d$) \cite{Boos:2020ccj}:
\begin{align}
\label{eq:gf-ultrarelativistic-simplification}
\lim\limits_{\gamma\rightarrow\infty} G_d(\bar{r}) = \frac{1}{\sqrt{2}} G_{d-1}(r_\perp)\delta(u-u') \, .
\end{align}
This collapses the integrals in the final solution and introduces the lower-dimensional Green functions $G_2$ and $G_4$, respectively:
\begin{align}
\dd s^2 &= -2\dd u \dd v + \phi \dd u^2 + 2 \ts{A} \dd u + \dd \ts{x}_\perp^2 \, , \\
\phi &= 2\sqrt{2} \lambda(u) G_2(r_\perp) \, , \\
\ts{A} &= 2\pi\kappa j{}_{ij} x{}_\perp^i \dd x{}_\perp^j G_4(r_\perp) \, .
\end{align}
Rewriting this in cylindrical coordinates one finds
\begin{align}
\phi(u,\rho) &= -\frac{\sqrt{2}\kappa\lambda(u)}{2\pi} \log \rho \, , \\
\ts{A}(u) &= \frac{\kappa j(u)}{2\pi} \dd\varphi \, .
\end{align}
This gravitational field depends on the outgoing null coordinate $u$, but allows the definition of a topological invariant of each time slice via
\begin{align}
j(u) = \frac{1}{8\pi G} \oint_{C_\rho} \ts{A} = \int\limits_0^{2\pi} \frac{j(u)}{2\pi} \dd\varphi \, .
\end{align}
In the non-singular field theory, Eq.\eqref{eq:gf-ultrarelativistic-simplification} is still valid and we hence obtain at the very end
\begin{align}
\ts{A}_\text{reg}(u) &= \frac{\kappa j(u)}{2\pi} \Delta_4(r_\perp) \dd\varphi \, .
\end{align}
This gives again a now radius-dependent magnetization,
\begin{align}
j_\text{reg}(u, r_\perp) = \frac{1}{8\pi G} \oint_{C_\rho} \ts{A}_\text{reg} = j(u) \Delta_4(r_\perp) \, ,
\end{align}
in strict analogy to the solenoid case and rotating string case, but in a time-dependent setting.

\subsection{Intermediate summary}

\begin{figure}[!htb]
    \centering
    \includegraphics[width=0.45\textwidth]{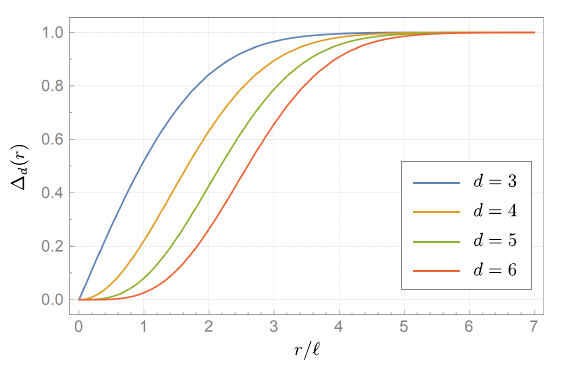}
    \caption{We plot the deviation functions $\Delta_d(r)$ for various values of $d$. For the most part (with the exception of $d=4$) they interpolate smoothly between $0$ and $1$.}
    \label{fig:delta}
\end{figure}

So far, we have seen that many topological invariants lose that meaning in non-singular field theories, and we managed to express that deviation via the functions $\Delta_d(r)$ for various choices of $d$ (depending on the cohomology class of the topological charge). For illustrative purposes, we visualize these deviation functions $\Delta_d(r)$ for the values $d=3,4,5,6$ in Fig.~\ref{fig:delta} for the toy model of Eq.~\eqref{eq:f-choice}. As expected, the functions interpolate between $0$ and $1$ on a characteristic length scale of $\ell$, corresponding to the UV regulator.

\section{Extended sources and the shell theorem in gravity}
\label{sec:extended}

\begin{figure}[!htb]
    \centering
    \includegraphics[width=0.45\textwidth]{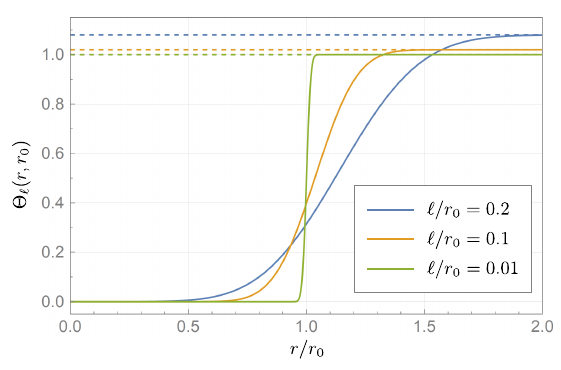}
    \caption{We plot the function $\Theta_\ell(r,r_0)$ for various values of $\ell$ expressed in units of $r_0$. Generally speaking, this function varies the most in the region where $|r-r_0|/\ell \sim 1$. Perhaps somewhat counter-intuitively, this function does not asymptote to unity, but rather to an $\ell$-dependent value. We visualize these different asymptotic values with dashed lines.}
    \label{fig:theta}
\end{figure}

Before we delve into nonlinear theory, let us address the case of extended spherically symmetric sources in Newtonian gravity, and, later, its non-singular generalization. The Poisson equation of Newtonian gravity is
\begin{align}
\nabla^2 \phi = 4\pi G \rho \, ,
\end{align}
which implies the shell theorem for spherically symmetric mass distributions:
\begin{itemize}
\item[(i)] The gravitational field outside a spherically symmetric mass distribution is given by the inverse-square low featuring the total mass contained inside that sphere.\\[-1.4\baselineskip]
\item[(ii)] The gravitational field inside a hollow sphere with spherically symmetric mass distribution vanishes.
\end{itemize}
Similar theorems hold in field theories with inverse-square laws, such as electrostatics. However, if one instead considers a non-singular field theory governed by the equation
\begin{align}
f(\ell^2\nabla^2) \nabla^2 \phi = 4\pi G \rho \, ,
\end{align}
the above theorem loses its validity due to the departure of the $1/r^2$-behavior of the gravitational force. At the level of the Green functions, this can be illustrated by
\begin{align}
\nabla^2 \mathcal{G}_d(r) &= - \, f^{-1}(\ell^2 \nabla^2) \, \delta{}^{(d)}(\ts{x}) \equiv - \, \delta{}^{(d)}_\ell(\ts{x}) \, , \tag{\ref*{eq:gf-f-2}}
\end{align}
where $\delta{}^{(d)}_\ell(\ts{x})$ denotes a nascent $\delta$-function that reduces to the distributional one in the limit of $\ell\rightarrow 0$. For example, Eq.~\eqref{eq:f-choice} implies (see also appendix \ref{app:gf})
\begin{align}
\label{eq:nascent-delta}
\delta{}^{(d)}_\ell(\ts{x}) = \frac{1}{(4\pi\ell^2)^{d/2}} e^{-\ts{x}^2/(4\ell^2)} \, .
\end{align}
Applied to the spherically symmetric case, a slight adjustment has to be made since a radial variable only assumes positive values, and one obtains
\begin{align}
\begin{split}
\delta{}^{(1)}(r-r_0) &= \frac{c(r_0,\ell)}{\sqrt{4\pi}\ell} e^{-(r-r_0)^2/(4\ell^2)} \, , \\
c(r_0,\ell) &= 2 \left[ 1 + \text{erf}\left( \frac{r_0}{2\ell} \right) \right]^{-1} \, .
\end{split}
\end{align}
Normalization issues aside, we note that in non-singular field theories featuring form factors $f(\ell^2\nabla^2)$ we can utilize standard Green function methods applied to an effective matter density
\begin{align}
\rho_\ell(\ts{x}) = \int\dd^d x' \, \delta{}^{(d)}_\ell(\ts{x}-\ts{x}') \, \rho(\ts{x}') \, .
\end{align}
This implies that even if a matter distribution has no support in a certain region (i.e.~$\rho(\ts{x}) = 0$) one may still have $\rho_\ell(\ts{x})\not=0$ in that same region. Putting this to the test, let us consider the mass density of an infinitesimally thin shell of mass $M$ and radius $r_0$,
\begin{align}
\rho(r) = \frac{M}{4\pi r_0^2} \delta{}^{(1)}(r-r_0) \, .
\end{align}
The effective density is hence
\begin{align}
\rho_\ell(r) &= \frac{M}{4\pi r_0^2} \frac{c(r_0,\ell)}{\sqrt{4\pi}\ell} e^{-(r-r_0)^2/(4\ell^2)} \, .
\end{align}
We then have for the gravitational force per unit mass
\begin{align}
\frac{1}{m} F_\text{N}(r) &= - \frac{G M}{r^2} \Theta_\ell(r,r_0) \, , \\
\Theta_\ell(r,r_0) &= \frac{c(r_0,\ell)}{\sqrt{4\pi}\ell r_0^2} \int\limits_0^r \dd \bar{r} \, \bar{r}^2 e^{-(\bar{r}-r_0)^2/(4\ell^2)} \, .
\end{align}
Explicit evaluation gives
\begin{align}
\begin{split}
\frac{\Theta_\ell(r,r_0)}{c(r_0,\ell)} &= \frac{r_0^2+ 2\ell^2}{2r_0^2}\left[ \text{erf}\left( \frac{r-r_0}{2\ell} \right) + \text{erf}\left( \frac{r_0}{2\ell} \right) \right] \\
&\hspace{11pt}+ \frac{\ell}{\sqrt{\pi}r_0} e^{-r_0^2/(4\ell^2)} \\
&\hspace{11pt}- \frac{\ell}{\sqrt{\pi}r_0} \left( 1 + \frac{r}{r_0} \right) e^{-(r-r_0)^2/(4\ell^2)} \, ,
\end{split}
\end{align}
where we factored our the normalization $c(r_0,\ell)$ to the left-hand side for notational brevity. First, we note the following property:
\begin{align}
\lim\limits_{\ell\rightarrow 0} \Theta_\ell(r,r_0) &= \begin{cases} 1 ~ : ~ r > r_0 \\ 0 ~ : ~ r < r_0  \end{cases} \, .
\end{align}
In the limit of vanishing of $\ell$ we recover standard, singular field theory, and it is a simple statement of the shell theorem: inside the hollow shell, the gravitational force vanishes. However, the case of $\ell > 0$ is more interesting. Namely, in the limit of large distances, $r \gg r_0$, we do not simply recover $\Theta_\ell = 1$, but rather
\begin{align}
\lim\limits_{r\rightarrow\infty} \Theta_\ell(r,r_0) &= \left( 1 + \frac{2\ell^2}{r_0^2} \right) \frac{1 + \text{erf}\left( \frac{r_0}{2\ell} \right)}{2} + \frac{\ell e^{-r_0^2/(4\ell^2)}}{\sqrt{\pi}r_0}  \nonumber \\
&\approx 1 + \frac{2\ell^2}{r_0^2} + \text{exp. suppressed in } \ell \, .
\end{align}
See Fig.~\ref{fig:theta} for a graphical representation of the function $\Theta_\ell(r,r_0)$ for different values of the UV regulator $\ell$.

The gravitational force, in that setting, is given by
\begin{align}
\frac{1}{m} F_\text{N}(r \gg \ell) &= - \frac{G M}{r^2} \left( 1 + \frac{2\ell^2}{r_0^2} \right) \, ,
\end{align}
which corresponds to rescaling of the mass $M$ to an effective mass
\begin{align}
M_\text{eff} = M \left( 1 + \frac{2\ell^2}{r_0^2} \right) \, .
\end{align}
Comparing notes with Sec.~\ref{sec:linearized-gravity}, this dressing effect is absent from point sources in and seemingly only affects extended objects.

\section{Electro-vacuum black hole solutions in general relativity}
Based on our previous considerations, it is clear that the area of 2-spheres must scale differently than $r^2$ at small distances to account for the improved short-distance behavior of point sources we seek to study. At the same time, this different scaling can then preserve the notion of a topologically invariant charge $Q$. Let us hence explore the notion of topological invariants in the context of black holes in general relativity coupled to electromagnetism, wherein the curvature of spacetime may allow for such a deformation of surface elements. In order to automatically guarantee the proper scaling, we imagine the non-singular field strength of a point charge scale with the radius $r$ via the function $F(r) > 0$. Then, we make the following ``$Q$-preserving'' ansatz:
\begin{align}
\begin{split}
\label{eq:bcf-ansatz}
\boldsymbol{g} &= -B(r)\dd t^2 + C(r)\dd r^2 + \frac{1}{F(r)} \dd\Omega^2 \, , \\
\dd\Omega^2 &= \dd\theta^2 + \sin^2\theta \, \dd\varphi^2 \, , \\
\sqrt{-g} &= \frac{\sqrt{BC}\sin\theta}{F} \, , \\
\boldsymbol{F} &= \frac{Q}{4\pi\epsilon_0} F(r) \, \dd t \wedge \dd r \, .
\end{split}
\end{align}
This ansatz guarantees sure that the charge contained in a 2-sphere of radius $r$ is independent of $F$,
\begin{align}
\begin{split}
\label{eq:charge-0}
& \epsilon_0 \oint_{S^2_r} \star \boldsymbol{F} = \oint_{S^2_r} \frac{Q}{4\pi\epsilon_0} F \, \star \left( \dd t \wedge \dd r \right) \\
&= \epsilon_0 \oint_{S^2_r} \frac{Q}{4\pi\epsilon_0} F \, \frac{\sqrt{BC}\sin\theta}{F} \dd\theta \wedge \dd\varphi \\
&= Q \, \sqrt{BC} \, .
\end{split}
\end{align}
We will see shortly that $BC$ has to be set to unity in accordance with the electromagnetic field equations, which then fully implements the invariant notion of the charge $Q$ as a topological invariant even if the field $F(r)$ does \emph{not} scale as $1/r^2$ for small distances. To arrive there, the Maxwell equations take the form $\text{d}\boldsymbol{F} = 0$ (satisfied by the ansatz) as well as (we assume $BC > 0$)
\begin{align}
\partial{}_r \left( \frac{1}{\sqrt{BC}} \right) = 0 \, ,
\end{align}
implying that $BC$ is constant, which can be set to $1$ by rescaling the time coordinate. Now Eq.~\eqref{eq:charge-0} is more transparent,
\begin{align}
\label{eq:charge-1}
& \epsilon_0 \oint_{S^2_r} \star \boldsymbol{F} = Q \, .
\end{align}
Hence, the strength of the ansatz \eqref{eq:bcf-ansatz} becomes apparent since it guarantees both the notion of an invariant $Q$ as well as a rather simple road to solve the Maxwell equations. To understand the physical properties of this (for now, off-shell) metric more, the Komar mass $M_\text{K}$ of this static geometry can be derived using the Killing 1-form $\boldsymbol{\xi} = -B \,\dd t$ via
\begin{align}
\begin{split}
G M_\text{K}(r) &= \frac{1}{8\pi} \oint_{S^2_r} \star \dd \boldsymbol{\xi} = \frac{1}{8\pi} \oint_{S^2_r} B' \star \left( \dd t \wedge \dd r \right) \\
&= \frac{1}{8\pi} \oint_{S^2_r} \frac{B'}{F} \sin\theta \dd\theta \wedge \dd\varphi = \frac{B'}{2F} \, .
\end{split}
\end{align}
The energy-momentum tensor of $\boldsymbol{F}$ is given by
\begin{align}
\begin{split}
\boldsymbol{T} &= \epsilon_0 \left( F{}_{\mu\alpha}F{}_{\nu\beta}g{}^{\alpha\beta} - \frac14 g{}_{\mu\nu} F{}_{\alpha\beta}F{}^{\alpha\beta} \right) \dd x{}^\mu \dd x{}^\nu \\
&= \frac{Q^2}{32\pi^2\epsilon_0}\frac{F^2}{BC} \left[ B\,\dd t^2 - C\,\dd r^2 + \frac{1}{F} \dd\Omega^2 \right] \, .
\end{split}
\end{align}
Let us now substitute into the field equations
\begin{align}
R{}_{\mu\nu} - \frac12 R g{}_{\mu\nu} = 8\pi G \, T{}_{\mu\nu}
\end{align}
to obtain the three independent equations
\begin{align}
\begin{split}
& F'' - \frac32\frac{F'^2}{F} = 0 \, , \\
& B'' - \frac{B'F'}{F}-2q^2F^2 = 0 \, , \\
& \frac{B'F'}{2F^3} - \frac{B F'^2}{4F^4} + \frac{1}{F} = q^2 \, ,
\end{split}
\end{align}
where we defined $q^2 \equiv G Q^2/(4\pi\epsilon_0)$ for convenience. The equation for $F$ is solved immediately, and the remainder then gives a solution for $B$:
\begin{align}
\begin{split}
F(r) &= \frac{1}{(r+c_1)^2} \, , \\
B(r) &= 1 + \frac{c_2}{r+c_1} + \frac{q^2}{(r+c_1)^2} \, .
\end{split}
\end{align}
The function $B(r)$ has the following large-$r$ expansion:
\begin{align}
B = 1 + \frac{c_2}{r} + \frac{q^2 - c_1 c_2}{r^2} + \mathcal{O}\left(\frac{1}{r^3}\right) \, .
\end{align}
We shall denote $c_1 = \ell$ and we read off $c_2 = -2GM$ from the asymptotic expansion for large $r$. As such, we parametrize the solution as follows
\begin{align}
\begin{split}
\label{eq:final}
\dd s^2 &= -B(r) \, \dd t^2 + \frac{\dd r^2}{B(r)} + (r + \ell)^2\dd\Omega^2 \, , \\
\ts{F} &= \frac{Q}{4\pi\epsilon_0} F(r) \, \\
F(r) &= \frac{1}{(r+\ell)^2} \, , \\
B(r) &= 1 - \frac{2GM}{r+\ell} + \frac{q^2}{(r+\ell)^2} \, , \quad q^2 = \frac{GQ^2}{4\pi\epsilon_0} \, .
\end{split}
\end{align}
This corresponds to the Reissner--Nordstr\"om solution of general relativity with a radial variable shifted by $\ell$.

Since the Maxwell equations of curved spacetime are diffeomorphism invariant, it is straightforward to obtain the above metric by the coordinate transformation $r \rightarrow \tilde{r} = r + \ell$ from the Reissner--Nordstr\"om metric of general relativity. In fact, by allowing $r \in [-\ell, \infty)$ we recover the Reissner--Nordstr\"om manifold identically. However, when written in the conventional form, the relationship between the $1/r^2$-falloff of the Maxwell field strength and the $r^2\dd\Omega^2$ in the spherical part of the metric, in relation to the notion of topological invariants, is somewhat obfuscated. Hence, the true strength of the ``Q-perserving'' ansatz \eqref{eq:bcf-ansatz} lies \emph{not} in generating a non-singular solution of the Einstein field equations. Rather, it demonstrates that a non-singular Maxwell field necessitates the deformation of 2-spheres if the notion of its topological invariant is to remain unchanged. It is precisely this rescaling of the spherical part of the metric that is possible in the non-linear theory that marks an important departure from the previously encountered results that non-singular field theories necessarily have to feature altered topological quantities.

That being said, let us now briefly explore the properties of the metric \eqref{eq:final} for $\ell > 0$. First and foremost, as $r > 0$ in our considerations, this metric is finite in its entire coordinate range. This can be verified, for example, by the curvature invariants
\begin{align}
C{}_{\mu\nu\rho\sigma}^2 = \frac{48[GM(r+\ell)-q^2]^2}{(r+\ell)^8} \, , ~ R{}_{\mu\nu}^2 = \frac{4 q^4}{(r+\ell)^8} \, .
\end{align}
In order to simplify the subsequent discussions, let us express all dimensionful quantities in units of $GM$,
\begin{align}
\hat{r} = \frac{r}{GM} \, , \quad \hat{\ell} = \frac{\ell}{GM} \, , \quad \hat{q} = \frac{q}{GM} \, .
\end{align}
This metric has two horizons located at
\begin{align}
\hat{r}_\pm = 1 - \hat{\ell} \pm \sqrt{1 - \hat{q}^2} \, .
\end{align}
For this metric to describe a black hole, we require the existence of an outer horizon. In that case, besides $\hat{q} < 1$ as a necessary condition, we need to impose additionally
\begin{align}
\hat{\ell} < \hat{\ell}_\text{max} = 1 + \sqrt{1 - \hat{q}^2} \, .
\end{align}
The inner horizon vanishes if $r_- < 0$ such that
\begin{align}
\hat{\ell} > \hat{\ell}_\text{min} = 1 - \sqrt{1 - \hat{q}^2} \, .
\end{align}
Note that the metric is not differentiable at $r=0$ since the metric function $B(r)$ features a linear term in its small-$r$ expansion,
\begin{align}
\begin{split}
B(r) &= B(0) - \frac{2(\hat{q}^2-\hat{\ell})}{\hat{\ell}^3} r + \mathcal{O}\left(r^2\right) \, , \\
B(0) &= 1 - \frac{2}{\hat{\ell}} + \frac{\hat{q}^2}{\hat{\ell}^2} \, .
\end{split}
\end{align}
Additionally, $B(0) \not=1$ for generic choices of $\ell$ (given parameters $q$ and $M$). The choice in the parameter $\ell$ can be parametrized as
\begin{align}
\hat{\ell} = \hat{\ell}(\alpha) = \frac{\hat{q}^2}{\alpha} \, ,
\end{align}
where $\alpha > 0$ is a positive number whose value will be determined in two scenarios.

\subsection{Case 1: Vanishing linear term}
A geometry that is spherically symmetric and regular at $r=0$ needs to feature a vanishing Weyl tensor. For the above parametrization of $\ell$ one finds
\begin{align}
\hat{C}{}_{\mu\nu\rho\sigma}^2(r=0) = \frac{48\alpha^6(1-\alpha)^2}{\hat{q}^{12}} \, ,
\end{align}
which vanishes if $\alpha=1$. For this choice, the linear term in the short-distance behavior of $B(r)$ cancels and
\begin{align}
B(\hat{r}) = 1 - \frac{1}{\hat{q}^2} + \frac{\hat{r}^2}{\hat{q}^6} + \mathcal{O}(\hat{r}^3) \, .
\end{align}
However, since $\hat{q} < 1$ in order to have a black hole, one immediately finds that in this scenario $B(0) < 0$. Also, the inner horizon vanishes since for $\alpha=1$ and $\hat{q}^2 < 1$
\begin{align}
\hat{\ell}(1) = \hat{q}^2 > 1 - \sqrt{1 - \hat{q}^2} \, ,
\end{align}
which can potentially avoid the mass inflation problem \cite{Poisson:1989zz}. We are left with a non-singular black hole electro-vacuum solution of electric charge $Q$ and ADM mass $M$. However, $B(0) < 0$ implies that the 3-volume picks up this unusual normalization factor via $\dd r/\sqrt{B(r)}$ close to $r = 0$, leading to a solid angle defect.

\subsection{Case 2: Proper normalization}
Demanding instead $B(0) = 1$ leads to $\alpha = 2$. The resulting metric is now properly normalized at the origin, without solid angle defects, but it features an inner horizon for all $\hat{q} \leq 1$ since for $\alpha = 2$ 
\begin{align}
\hat{\ell}(2) = \frac{\hat{q}^2}{2} < 1 - \sqrt{1 - \hat{q}^2} \, .
\end{align}
However, the metric---even though finite---is not differentiable at $r=0$ and hence not regular in that sense. At small distances, the metric scales as
\begin{align}
B(r) = 1 - \frac{8 \hat{r}}{\hat{q}^4} + \mathcal{O}(r^2) \, .
\end{align}
Even though we have verified that the invariant $[\nabla_\lambda C{}_{\mu\nu\rho\sigma}]^2$ is finite at $r=0$, it is expected that invariants of higher order involving derivatives of the curvature tensor will eventually diverge at the origin.

\subsection{General comments}

\begin{figure}[!htb]
    \centering
    \includegraphics[width=0.45\textwidth]{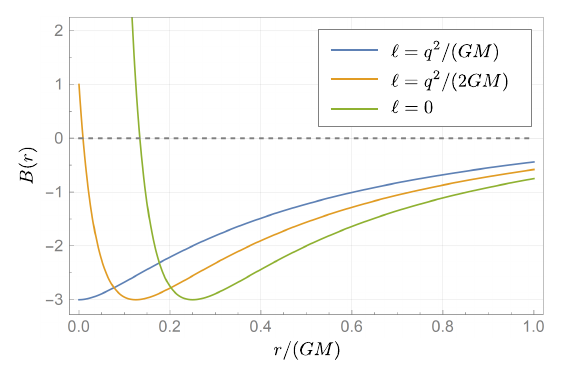}
    \caption{We plot the metric function $B(r)$ for $q=0.5 GM$ and three choices of $\ell$. In the first case, $\ell=q^2/(GM)$, we see that the metric function is regular at $r=0$ but remains negative. In the second case, $\ell = q^2/(2GM)$, the metric function approaches unity, $B(0) = 1$, but does so with a non-vanishing slope. The third line shows the Reissner--Nordstr\"om line element ($\ell = 0$) for reference. At large distances $r \gg GM$, all three metric functions converge. Note that in the first case there exists an inner horizon, like in the Reissner--Nordstr\"om metric, but in the second case the inner horizon is absent.}
    \label{fig:b}
\end{figure}

The choice of the regulator $\ell$ is hence intimately related to the short-distance properties of the metric; see Fig.~\ref{fig:b} for a graphical representation of the metric function $B(r)$ for cases 1 and 2 in comparison to the Reissner--Nordstr\"om metric function. It should be mentioned that the spacetime nature of the location $r=0$ is strictly distinct: since the case $\alpha=1$ has no inner horizon, $r=0$ is a spacelike surface, whereas for $\alpha=2$ it is timelike.

In that context, we would like to mention that for general $\ell$ including the region $r<0$ reinstates the singularity problem of the metric, and it is intimately tied to the issue of geodesic completeness under analytic continuation \cite{Carballo-Rubio:2019fnb}. For general $\ell$, therefore, the metric presented in this section cannot be considered geodesically complete. However, several well-known non-singular black hole models are not geodesically complete \cite{Zhou:2022yio}. The choice $\alpha = 1$ [$\ell = q^2/(GM)$] removes the linear term at the origin and introduces a purely quadratic $r$-dependence, making the metric symmetric under $r \rightarrow -r$ at the origin, which then reinstates the geodesic completeness.

Taking a closer look at the regulator $\ell$ (here for the choice $\alpha=1$) we arrive at the magnitude
\begin{align}
\ell & \sim \frac{1}{4\pi\epsilon_0 c^2} \frac{Q^2}{M} = \frac{(Q[\text{e}])^2}{M[\text{MeV}]} \times 0.144 \, \text{fm} \, .
\end{align}
Interestingly, the dependence on the gravitational constant $G$ drops out. For astrophysical objects of negligible charge this length scale is miniscule. However, since for astrophysical effects $\hat{q} \ll 1$, the incorrect normalization of the metric at $r=0$ in the case of $\alpha=1$ is a huge effect, since $|1-1/\hat{q}^2| \gg 1$. In the case of $\alpha=2$, however, we also find that the linear term in the short-distance limit scales with $1/\hat{q}^3 \gg 1$. Hence, in either case, close to $r=0$ one expects major deviations from a regular geometry.

For elementary particle masses and charges, however, the regulator is comparable to their typical size. It should be noted, though, that in such a case $\hat{q}^2 \gg 1$ such that these objects do not feature a horizon. In this case, the usual problem of naked singularities as encountered in general relativity does not occur, since the metric is manifestly finite.

Moreover, note that this non-singular black hole configuration does not have a mass gap \cite{Frolov:2015bta}: most non-singular black hole models (see below), as well as field configurations described in modified gravity theories, only describe a black hole when the regulator length $\ell$ is bound (roughly) by the mass of the gravitating object, $\ell \lesssim GM$. In the present case, no such limitation exists, likely because the regulator encountered here is given by the fundamental charge of the object, and is therefore not a free parameter that may become arbitrarily large compared to the black hole geometry, since $q \lesssim GM$ for astrophysical objects.

Finally, we note that the coordinate transformation $r \rightarrow r + \ell$ can be employed to the Kerr--Newman metric with cosmological constant to arrive at a similarly finite metric with a rotation parameter.

\section{Conclusions and discussion}

The singularity problem of classical field theories has long motivated the study of modified theories that do not feature such singularities. This either happens purely in the classical regime, where such theories are regarded as approximative classical limits of an underlying finite quantum field theory, or at the full quantum level, where such theories can lead to improved ultraviolet behavior in the calculation of amplitudes. At the end of the day, however, it is of paramount importance to test these modified theories with observations.

It is a common phenomenological problem that effects of UV completions of field theories are miniscule in physical observables. A prominent recent example is that of black hole shadows, where theoretically modified gravity theories lead to different shadow regions, but are in practice not detectable by a good margin \cite{Vagnozzi:2022moj}. It is hence of interest to search for so-called ``smoking gun'' signatures of modified field theory. In this paper, we have identified a promising direction where to search for such a possibility in the theoretical realm.

Namely, the notion of a topological invariant depends sensitively on the cohomology class of spacetime: what type of loops (or hypersurfaces) are contractible to a point, and which ones are not? For example, non-trivial winding numbers can only be defined in three dimensional space if that space has, say, the $z$-axis removed. On the physical side, in linear theories, removed spacetime domains correspond to the location of point-like sources described by $\delta$-function shaped matter densities. In non-linear theories, singularities themselves shape the cohomology class of spacetime. For this reason, the topological properties of spacetimes (and the properties of their topological invariants, by extension) hence allow us to make statements about their matter content, and, in particular, the regularity of their matter fields.

To that end, in the linear theory we have considered a wide class of non-singular field theories described by UV-improved Green functions $\mathcal{G}_d = G_d(r)\Delta_d(r)$, where $G_d(r)$ denotes the Green function from standard, singular field theory, and $\Delta_d(r)$ denotes the deviation. Then, we extracted various topological invariants from physical examples ranging from electrodynamics to weak-field gravity. Broadly speaking, a topological invariant $Q$ in the standard field theory is demoted to a spacetime-dependent quantity,
\begin{align}
Q ~ \rightarrow ~ Q(r) = Q \, \Delta_d(r) \, ,
\end{align}
where $d$ depends on the topological properties of the source in standard, singular field theory. Intuitively, this happens because non-singular field theories turn distributional sources supported by $\delta$-functions into smooth functions with infinite support. In most cases, however, UV-improved theories come with a regulator length scale $\ell > 0$, and for distances $r \gg \ell$ one recovers the results of standard field theory, such that the notion of a topological invariant is restored asymptotically,
\begin{align}
Q(r\rightarrow\infty) \approx Q \, .
\end{align}
In the linear theory for extended sources we then applied this concept to a thin sphere and demonstrated the departure from the well-known shell theorem that states, inter alia, that the gravitational force inside a spherically symmetric hollow sphere vanishes: in non-singular field theories, there is a gravitational field---mutatis mutandis for other field theories with an inverse-square law. Moreover, we showed that the asymptotic mass of such an extended mass distribution may pick up a dressing in form of a multiplicative factor that is close to unity.

Last, we constructed a metric ansatz within general relativity that allows for a non-singular electromagnetic field of a point particle, but has a radial sector that automatically guarantees that the area of spheres scales with the inverse of that field strength. This restores the notion of a topological invariant for the electromagnetic field, and is only possible in curved spacetime. Solving the field equations, we showed that this metric coincides with the Reissner--Nordstr\"om metric of a charged black hole, wherein the radial variable $r$ has been shifted to positive values only, $r \rightarrow r + \ell$. The linear and quadratic curvature invariants of this metric are finite, but depending on the choice of $\ell$, curvature invariants involving derivatives may still diverge at $r=0$. A summary of our main results can be found in Table~\ref{tab:main-results}.

In the remainder, we would like to conclude by connecting the ideas presented in this paper to other related research directions.

\subsection{Non-singular black holes}

There has been much activity in the field of non-singular black holes, and we would like to close by drawing some comparisons. The field can be divided into a few categories. First of all, there are non-singular black hole models that do not solve the field equations of general relativity, but whose form is postulated by various physical considerations \cite{Bardeen:1968,Dymnikova:1992ux,Hayward:2005gi,Frolov:2016pav}.\footnote{We would like to draw some attention to a recently proposed model by the author wherein the regulator length scale is macroscopic size, with potentially percent-level effects at the black hole horizon scale \cite{Boos:2023icv}.} On the other hand, it is also possible to solve particular field equations encountered in non-linear electrodynamics coupled to general relativity and show that the resulting metrics are bona fide non-singular black hole solutions \cite{Ayon-Beato:1998hmi,Ayon-Beato:1999kuh,Dymnikova:2004zc,Bronnikov:2000vy}. In that context, we would also like to point out recent work by Bueno \textit{et al.} who considered general relativity coupled to higher-order curvature corrections and constructed exact, non-singular black hole solutions in five and higher dimensions \cite{Bueno:2024dgm}. Third, practitioners of various approaches to quantum gravity have constructed hybrid models that are inspired by the corresponding underlying theory of quantum gravity but are in a suitable sense effective field theory limits thereof. A prominent example is asymptotic safety \cite{Reuter:1996cp}, wherein so-called renormalization group improved black holes have been considered \cite{Bonanno:2000ep,Platania:2019kyx,Bosma:2019aiu,Boos:2023xoq}.

Let us mention that in the context of non-singular, renormalization group-improved black holes \cite{Bonanno:2000ep}, a running gravitational coupling $G(k)$ can be read off by defining $G(r) = G_0 \, r/(r+\ell)$ and then employing a cutoff identification $k = k(r)$. For $k(r) = 1/\sqrt{G_0 M r}$ (which is one out of many possible choices), we find
\begin{align}
G(k) = \frac{G_0}{1 + \omega G_0 k^2} \, , \quad \omega = \frac{Q^2}{4\pi\epsilon_0} \, ,
\end{align}
whose functional form coincides with that encountered in quantum Einstein gravity \cite{Reuter:1996cp}. Since $\omega$ is related to a putative ultraviolet fixed point of the gravitational coupling, the above links our regular metric to the resolution of singularities within asymptotic safety \cite{Bosma:2019aiu}.

Other points of contact with the literature on non-singular black holes exist with the work of Klinkhamer \cite{Klinkhamer:2013wla}, who constructed a Schwarzschild-type black hole in general relativity with a central part topologically removed; with the Simpson--Visser metric \cite{Simpson:2018tsi} (which features a non-standard area element and can hence also be thought of describing a wormhole) as well as the author's previous work \cite{Boos:2021kqe}, which was motivated by defining a radial distance by the inverse of a regular potential (unlike in this work, where we defined the area by the inverse of the regular  field strength).

\subsection{Dressing of observables}

It has been argued that the construction of diffeomorphism-invariant observables is possible in the context of quantum gravity if one endows them with an operational interpretation \cite{Donnelly:2015hta}. For example, it is only possible to define a diffeomorphism-invariant observable for an electron via its Wilson line if one simultaneously takes into account its gravitational field that is dragged along with the electric charge. While this example stems from an entirely different field, one cannot help but notice the operational similarity to the prescription proposed in this work: the electromagnetic field of an electric point charge can only be regular of one simultaneously modifies its gravitational field.

Moreover, the dressed total charge encountered for the hollow mass shell in Sec.~\ref{sec:extended} reminds of previous work by the author, where a Lippmann--Schwinger equation was utilized to demonstrate that far-distance observables (such as scattering amplitudes and vacuum fluctuations) can be sensitive to short-distance physics \cite{Boos:2018kir,Boos:2019fbu}. The distance-dependent topological invariants computed in this paper hence present a possible avenue to search for UV-IR connections in field theory.

\subsection{Action selection principles for non-singular black hole metrics}

Last, we want to point out that the final metric \eqref{eq:final}, even though finite for all $r > 0$, is not completely regular around $r=0$. Higher-order gravity theories, featuring not only quadratic curvature invariants in their actions but also derivatives thereof, should be sensitive to such field configurations. In fact, it has long been argued that curvature invariants play a decisive role in the dynamical selection of viable non-singular black hole geometries via their on-shell values in variational principles \cite{Giacchini:2021pmr,Knorr:2022kqp,Borissova:2023kzq,Borissova:2024hkc}. It is plausible that invariants such as $\mathcal{R}^p \Box^n \mathcal{R}^q$ diverge at $r=0$ for the metric \eqref{eq:final} for some choice of $p$, $q$, and $n > 0$, where $\mathcal{R}$ is a curvature expression. In that case, applying the ``Q-perserving'' ansatz \eqref{eq:bcf-ansatz} to field equations of higher-order gravity may lead to insightful and novel black hole solutions.

\subsection{Observational consequences}

\begin{figure}[!htb]
    \centering
    \includegraphics[width=0.45\textwidth]{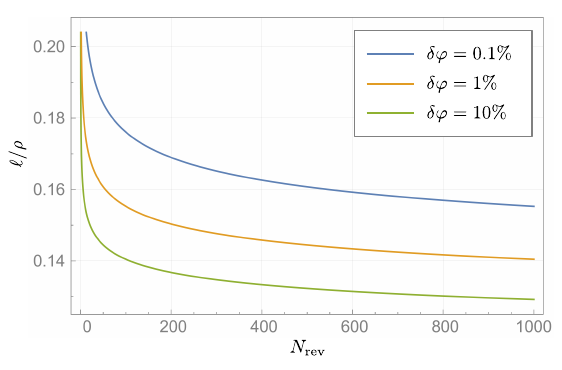}
    \caption{For a given relative Aharonov--Bohm phase deviation, we plot the relation of $\ell/\rho$ to the number of revolutions $N_\text{rev}$. For a fixed $\rho$ and $N_\text{rev}$, the above diagram shows the lower limits of the UV parameter $\ell > 0$.}
    \label{fig:obs}
\end{figure}

The key result for linear theories is the destruction of topological invariants by UV regularization. It is hence instructive to search for potentially observable consequences of such mechanisms. A first starting point is the Aharonov--Bohm effect that has been experimentally confirmed in the Abelian case of U(1) gauge theory \cite{Chambers:1960xlk,Tonomura:1982}, as well as non-Abelian gauge theory \cite{Yang:2019tem} and weak-field gravity \cite{Overstreet:2021hea}. Therein, typical phase shifts assume values of roughly $0.1\pi$ at typical length scales of micrometers. Since deviations from topological properties of the string cohomology class are parametrized by the function $\Delta_4(\rho)$, we can use these phase shifts to place constraints on the regulator scale $\ell$. Of course, the precise constraint depends on the type of non-singular field theory under consideration. In the present case, let us utilize
\begin{align}
\Delta_4(\rho) = 1 - \exp\left[-\rho^2/(4\ell^2)\right] \, .
\end{align}
Demanding a 1\% effect, $1-\Delta_4(\rho=\mu\text{m}) \lesssim 0.01 \times 0.1\pi$, then results in
\begin{align}
\ell \lesssim 0.21 \mu\text{m} \, .
\end{align}
By considering instead topological phases for multiple revolutions (say, where electrons circle a solenoid field thousands of times) this constraint can be successively improved since the relative phase shift per revolution stays the same but the cumulative phase shift $\delta\varphi$ will scale with the number of revolutions $N_\text{rev}$ such that
\begin{align}
\ell > \frac{r}{4\sqrt{\log{(N_\text{rev}/\delta\varphi)}}} \, .
\end{align}
In Fig.~\ref{fig:obs} we plot the lower limits for $\ell$ given a desired phase shift strength $\delta\varphi$ as a function of revolutions $N_\text{rev}$. While these considerations depend on the precise model of non-singular field theory, a generic prediction is the radius-dependence of the Aharonov--Bohm phase.

\begin{table}
\begin{tabular}{lclclc} \hline \hline
Quantity && Expression && Type \\ \hline
 && && \\[-7pt]
Electric charge && $\displaystyle Q_\text{reg}(r) = Q \Delta_5(r)$ && 3D \\[7pt]
Magnetization && $\displaystyle \mu_\text{reg}(\rho) = \mu \, \Delta_4(\rho)$ && 2D \\[7pt]
Angle deficit && $\displaystyle \delta\varphi_\text{reg}(\rho) $ && 2D \\[7pt]
Angular momentum && $\displaystyle j_\text{s,\,reg}(\rho) = j_s \Delta_4(\rho) $ && 2D \\[7pt]
Komar mass && $\displaystyle G M_\text{K,\,reg}(r) = GM \Delta_5(r)$ && 3D \\[7pt] \hline
 && && \\[-7pt]
Mass of hollow sphere && $\displaystyle M_\text{eff} \approx M \left(1 + \frac{2\ell^2}{r_0^2} \right)$ && 3D \\[7pt] \hline
 && && \\[-7pt]
Non-linear electric charge && $\displaystyle Q_\text{reg}(r) = Q$ && 3D 
\\[7pt] \hline\hline
\end{tabular}
\caption{In the linear theory of point-like particles (first group of rows), topological charges such as electric charge, magnetization, angular momentum, and Komar mass are modified via the deviation functions $\Delta_d(r)$. The angle deficit alone is given by an integral over the Green function $\mathcal{G}_2$ but is nevertheless a function of transverse distance to the string in non-singular field theories. In the linear field for a hollow sphere, we find an effective, radius-independent mass that persists to spatial infinity. And, last, in the context of general relativity, we can arrive at a well-defined, constant electric charge for a non-singular electric and finite gravitational field, at the cost of introducing topological defects close to $r=0$.}
\label{tab:main-results}
\end{table}

\section{Acknowledgements}

I thank two anonymous reviewers for their feedback, which helped to improve the clarity of the manuscript. It is my additional pleasure to thank Ivan Kol\'a\v{r} (Charles University) and Robie Hennigar (Durham University) for comments on an earlier version of this draft. I am grateful for support as a Fellow of the Young Investigator Group Preparation Program, funded jointly via the University of Excellence strategic fund at the Karlsruhe Institute of Technology (administered by the federal government of Germany) and the Ministry of Science, Research and Arts of Baden-W\"urttemberg (Germany).

\appendix

\section{Singularity-free Green functions}
\label{app:gf}
The Green functions $G_d(r)$ and $\mathcal{G}_d(r)$ in $d$ spatial dimensions are defined via
\begin{align}
\nabla^2 G_d(r) =  -\delta{}^{(d)}(\ts{x}) \, , \tag{\ref*{eq:gf}} \\
\nabla^2 f(\ell^2 \nabla^2) \mathcal{G}_d(r) =  -\delta{}^{(d)}(\ts{x}) \, , \tag{\ref*{eq:gf-f}} \, .
\end{align}
For the choice
\begin{align}
f(\ell^2\nabla^2) = \exp\left( -\ell^2\nabla^2 \right) \tag{\ref*{eq:f-choice}}
\end{align}
considered here for simplicity, they take the form
\begin{align}
G{}_2(r) &= -\frac{1}{2\pi}\log\left(\frac{r}{r_0}\right) \, , \\[5pt]
\mathcal{G}_2(r) &= -\frac{1}{2\pi} \left[ \log\left(\frac{r}{r_0}\right) - \frac12 \, \text{Ei}\left( -\frac{r^2}{4\ell^2} \right) \right] \, , \\[5pt]
G_3(r) &= \frac{1}{4\pi r} \, , \\[5pt]
\mathcal{G}_3(r) &= \frac{\text{erf}[r/(2\ell)]}{4\pi r} \, , \\[5pt]
G_4(r) &= \frac{1}{4\pi^2 r^2} \, , \\[5pt]
\mathcal{G}_4(r) &= \frac{1 - \exp\left[-r^2/(4\ell^2)\right]}{4\pi^2 r^2} \, , \\[5pt]
G_5(r) &= \frac{1}{8\pi^2 r^3} \, , \\[5pt]
\mathcal{G}_5(r) &= \frac{ \text{erf}[r/(2\ell)] - [r/(\sqrt{\pi}\ell)]\exp\left[-r^2/(4\ell^2)\right] }{8\pi^2 r^3} \, .
\end{align}
Here, we made use of the auxiliary definition
\begin{align}
\text{Ei}(-x) = -E_1(x) = - \int\limits_x^\infty \dd z \, \frac{e^{-z}}{z} \, , \quad x > 0 \, .
\end{align}
They satisfy the recursion formulae
\begin{align}
\partial_r G_d &= -2\pi r G_{d+2}(r) \, , \tag{\ref*{eq:gf-rec}} \\
\partial_r \mathcal{G}_d &= -2\pi r \mathcal{G}_{d+2}(r) \, . \tag{\ref*{eq:gf-rec-f}}
\end{align}
For Cartesian coordinates, the nascent $\delta$-function \eqref{eq:gf-f-2}, for the above choice of $F$ in Eq.~\eqref{eq:f-choice}, takes the form \cite{Boos:2021suz}
\begin{align}
\tag{\ref*{eq:nascent-delta}}
\delta{}^{(d)}_\ell(\ts{x}) = \frac{1}{(4\pi\ell^2)^{d/2}} \exp\left[ -\frac{\ts{x}^2}{4\ell^2} \right] \, ,
\end{align}
and it admits the product representation
\begin{align}
\delta{}^{(d)}_\ell(\ts{x}) = \prod\limits_{i=1}^d \frac{1}{\sqrt{4\pi}\ell} \exp\left[ -\frac{(x^i)^2}{4\ell^2} \right] \, .
\end{align}
In case of radial coordinates spanning only half the real axis, proper care has to be taken to utilize the correct normalization such that indeed
\begin{align}
\int \dd^d x \, \delta{}^{(d)}_\ell(\ts{x}) = 1
\end{align}
for any choice of $\ell$.

\pagebreak

\end{document}